\documentclass[]{interact}

\usepackage{appendix}
\usepackage{xcolor}

\usepackage{epstopdf}
\usepackage[caption=false]{subfig}

\usepackage[numbers,sort&compress]{natbib}
\bibpunct[, ]{[}{]}{,}{n}{,}{,}
\renewcommand\bibfont{\fontsize{10}{12}\selectfont}
\makeatletter
\def\NAT@def@citea{\def\@citea{\NAT@separator}}
\makeatother

\theoremstyle{plain}

\theoremstyle{definition}

\theoremstyle{remark}

\usepackage{lmodern}   
\usepackage[T1]{fontenc}

\begin{document}


\title{Differential Equation-Constrained Exponential-Type Local Polynomial Regression Under Model Misspecification}

\author{
\name{Chunlei Ge$^\ast$\textsuperscript{a} and W. John Braun\textsuperscript{a}\thanks{$^\ast$CONTACT W. John Braun. Email: john.braun@ubc.ca}}
\affil{\textsuperscript{a}The University of British Columbia Okanagan, 3333 University Way, Kelowna, BC, Canada}
}

\maketitle

\begin{abstract}
The issue of model misspecification is critical, yet it is often regarded as unavoidable in applied statistical modeling. Model misspecification can be mitigated by incorporating informative features and strengthening model formulations, such as through the integration of domain knowledge or structural constraints. In this paper, we propose a regression framework constrained by differential equations, which leverages first-order differential equations and adapts local polynomial regression techniques. Specifically, we focus on the local exponential growth model, characterized by an exponential-type differential equation. For this model, we examine the asymptotic biases and variances of kernel estimators constructed using Taylor polynomials of varying degrees. To evaluate model robustness, we conduct simulation studies comparing different estimators under two misspecification scenarios varying the levels of misspecification.

\end{abstract}

\begin{keywords}
Nonparametric regression; Local polynomial regression; Model misspecification; Differential equations. 
\end{keywords}

\section{Introduction}

Model misspecification is a fundamental concern in statistical modeling because it can produce biased parameter estimates, reduce estimation efficiency, distort predictions, and compromise the validity of hypothesis tests.

Statistical tools for detecting and diagnosing model misspecification include residual analysis, cross-validation, information criteria such as AIC and BIC, and specification tests such as the Ramsey Regression Equation Specification Error Test (RESET) \citep{ramsey1969tests}. In addition, \citet{cheng2018bias} developed simple bias-reduction methods for nonparametric and semiparametric regression, particularly useful when the assumed parametric model is misspecified.

We aim to mitigate the impact of model misspecification by introducing a robust nonparametric approach. In this paper, we present a differential equation-constrained regression method that was originally proposed by \citet{ding2014estimation} and extend the study on local exponential growth models in our previous work \citet{ge2025differential} to investigate its performance under model misspecification. \citet{ding2014estimation} focused primarily on parameter estimation for differential equations, whereas our interest lies in local parameter estimation, allowing for greater flexibility in handling misspecified models. This approach leverages structural constraints for improved robustness, and our analysis extends to simulation studies demonstrating estimator behavior under varying degrees of misspecification.

Local polynomial regression, as introduced by \citet{fan1996local}, is a highly effective nonparametric method owing to its flexibility and capacity for local adaptation. This estimator evolved from the kernel regression approach known as the Nadaraya-Watson estimator, which is also referred to as the local constant estimator (\citet{nadaraya1964estimating}; \citet{watson1964smooth}). In this paper, the proposed a bias reduction method that incorporates differential equation constraints into conventional local polynomial regression. The resulting DE-constrained local polynomial regression method extends the applicability of local polynomial techniques to a broad class of differential equations, encompassing both linear and nonlinear forms.

The remainder of the paper is organized as follows. Section 2 introduces the differential equation-constrained regression model. Section 3 details a specific subclass, the local exponential growth model, defined by an exponential-type differential equation. Section 4 presents the DE-constrained estimator. Section 5 discusses the asymptotic properties of DE-constrained estimation. Section 6 illustrates the effectiveness of the proposed approach through simulation studies. Section 7 contains discussion and concluding remarks. Proofs of the main results and additional simulation figures are provided in the Appendix.

\section{Differential Equation-constrained Regression Model}

Given $n$ independent observations on an explanatory variable  $x$ and
a response variable $y$, we consider the following DE-constrained regression model:

\begin{equation}
y_i = g(x_i) + \varepsilon_i 
\mbox{ where } g'(x) = F(x,g(x)), \ \ \ \  x_i \in [a,b], \ \ i = 1, 2, \ldots, n.  
\label{equ:general} 
\end{equation}
for some Lipschitz continuous function $F$ and uncorrelated, mean-zero errors $\varepsilon_i$.

We assume that the design points have been randomly sampled from an interval $[a, b]$ according to a probability density function $f(x)$, or they have been selected according to a fixed sampling design in that interval. We assume that $f(x)$ is continuously differentiable to the second order and has a bounded second derivative.

In this paper, we focus on a first-order differential equation-constrained model; specifically, we will apply an exponential-type differential equation to our model. The description will be in the next section.

\section{Local Exponential Growth Model}

In model (\ref{equ:general}), we assume that $F(x,g(x))=\lambda g(x)$, therefore, we have a local exponential growth model (\citet{ge2025differential}):

\begin{equation}
y_i = g(x_i) + \varepsilon_i \mbox{ where } g'(x) = \lambda g(x), \ \ \ \  x_i \in [a,b], \ \ i = 1, 2, \ldots, n.  
\label{eqn:exp} 
\end{equation}

The differential equation $g'(x)=\lambda g(x)$ has an explicit solution  
\begin{equation}\label{expnls} g(x) = g(a) e^{\lambda (x - a)}.  \end{equation}

The global parameter $\lambda$ can be estimated by nonlinear least-squares regression. In practice, we can apply the function \textbf{nls}() in the R Package \textbf{nlstools} (\citet{baty2015toolbox}) to the data.

\section{Differential Equation-constrained Regression Estimation}

The approach developed in this paper builds on the local polynomial regression framework of \citet{fan1996local} and extends the methodology beyond the differential equation–constrained models originally considered by \citet{ding2014estimation}. While \citet{ding2014estimation} focuses on the estimation of the parameters of the ordinary differential, we are interested in the estimation of local parameters.

we assume that the design density, kernel, and bandwidth satisfy the conditions outlined in \citet{fan1996local}. For a given evaluation point $x_0$, a DE-constrained local polynomial regression estimator for $g(x)$ is obtained
by minimizing the local least-squares objective function:
\begin{equation}\label{eqn:genmethod} \sum_{i=1}^n \{y_i - g_k^*(x_i)\}^2 K_h(x_i-x_0). \end{equation}
where 
\begin{equation}\label{eqn:star} g_k^*(x_i) = g(x_0) + (x_i-x_0) g'(x_0) + \cdots + \frac{(x_i-x_0)^k}{k!} g^{(k)}(x_0)  \end{equation}
is the $k^{th}$-degree Taylor approximation of $g(x_i)$ about $x_0$.  This procedure will give us the $k^{th}$-degree DE-constrained estimation. Specifically, 
to achieve the first-degree DE-constrained estimation, by incorporating information from the differential equation in model (\ref{eqn:exp}) , that is,
\[ g'(x) = \lambda g(x), \]
we have
\[ g_1^*(x_i) = g(x_0) + (x_i - x_0)\lambda g(x_0), \]

then the local least squares {eqn:genmethod} will be 

\[ 
\sum_{i=1}^n \{y_i - g_1^*(x_i)\}^2 K_h(x_i - x) 
= \sum_{i=1}^n \{ y_i - g(x_0) - \lambda g(x_0) (x_i - x)\}^2 K_h(x_i-x_0).  \]

Optimizing the expression on the right leads to the first-degree DE-constrained estimator,denoted as $\text{DE}_1$ estimator, 
\begin{equation}\label{eqn:DE1} \widehat{g}_1(x_0) = \frac{
\sum_{i=1}^n y_i(1+(x_i-x_0)\lambda)K_h(x_i-x_0)}{\sum_{i=1}^n (1 + (x_i-x_0)\lambda)^2 K_h(x_i-x_0)}. \end{equation}

Additional derivatives can be taken.  In general, by applying the $k^{th}$ degree Taylor expansion for $g(x_i)$ in a sufficiently small neighborhood of  $x_0$,we can obtain the $k^{th}$ degree DE-constrained estimator, $\text{DE}_p$ estimator (\citet{ge2025differential}): 

\begin{eqnarray}
\widehat{g}_k(x_0) 
&= \arg\min_{g(x_0)} \sum_{i=1}^n \left\{y_i- \sum_{p=0}^k \frac{1}{p!}(x_i-x_0)^p g^{(p)}(x_0)\right\}^2 K_h(x_i-x_0) \nonumber\\
&= \arg\min_{g(x_0)} \sum_{i=1}^n \left\{y_i-\sum_{p=0}^k \frac{1}{p!}(x_i-x_0)^p \lambda^{p}g(x_0)\right\}^2 K_h(x_i-x_0) \nonumber\\
&= \frac{\sum_{i=1}^n \{y_i \sum_{p=0}^k \frac{1}{p!}(x_i-x_0)^p \lambda^{p} \} K_h(x_i-x_0) }{\sum_{i=1}^n \{\sum_{p=0}^k \frac{1}{p!}(x_i-x_0)^p \lambda^{p} \}^2K_h(x_i-x_0)}.
\end{eqnarray}

\section{Asymptotic Properties}

When performing the conditional asymptotic analysis of the $k^{th}$ degree estimators, we make the following assumptions for model (\ref{eqn:exp}):

(I) $g(x_0)$, the mean function, has a bounded and continuous $k+1^{th}$ derivative in a neighborhood of  $x_0$. 

(II) $f(x_0)$, the design density, is twice continuously differentiable and positive.

(III) $K_h(x)$, the kernel function, is a nonnegative, symmetric and bounded PDF with compact support, on the interval $[-h, h]$. The kernel function satisfies $\int_{-\infty}^{\infty}K_h(w)dw=1$, $R(K)=\int K_1^2(w)dw < \infty$, and has finite moments up to sixth order.   We will also use the notation $\mu_{k}=\int w^{k}K_1(w)dw$, and $R_k = \int w^{k} K_1^2(w)dw$.  

(IV) The error variance $\sigma^2(x)$ is a smooth function on $[a, b]$.  

\subsection{Asymptotic Conditional Bias and Variance}

We summarize the results concerning the asymptotic conditional bias and variance of the DE-constrained estimators in the interior of the interval $[a,b]$ in the following two theorems. 

\textbf{Theorem 1 (Asymptotic Conditional Bias)} For regression model (\ref{eqn:exp}), under the assumptions (I) - (III), with $x_0 \in (a+h, b-h)$,  the $\text{DE}_k$ estimator, $\hat{g}_k (x_0)$, has asymptotic conditional bias

\begin{equation}
\mathrm{Bias}(\widehat{g}_k (x_0)|x_1,...,x_n) = \frac{1}{(k+1)!}\lambda^{(k+1)}g(x_0)h^{k+1}\mu_{k+1}+o_p(h^{k+1}),  \quad k \quad \text{odd},
\end{equation}
and when $k$ is even,
\begin{equation}
\mathrm{Bias}(\widehat{g}_k (x_0)|x_1,...,x_n) = \frac{1}{(k+1)!}\lambda^{(k+1)}g(x_0)\left(\frac{\lambda}{k+2}+\frac{f'(x_0)}{f(x_0)}\right)h^{k+2}h^{k+2}\mu_{k+2}+o_p(h^{k+2}).
\end{equation}

\textbf{Theorem 2 (Asymptotic Conditional Variance)} Under the assumptions (I) through (IV), with $x_0 \in (a+h, b-h)$,  the $\text{DE}_k$ estimator, $\hat{g}_k (x_0)$,  has asymptotic conditional variance
\begin{equation}
\mathrm{Var}(\widehat{g}_k (x_0)|x_1,...,x_n) =\frac{\sigma^2R(K)}{nhf(x_0)}+o_p\left(\frac{1}{nh}\right).
\end{equation}

From the results of Theorems 1 and 2, the differential equation-constrained local polynomial regression approach examined in this study effectively reduces asymptotic bias without substantially increasing variance.

The proofs of Theorems 1 and 2 are provided in the appendix.  

\subsection{Theoretical Comparison of Estimators}
In this section, we will compare the asymptotic properties of different estimators, assuming the correct model.
We will compare our DE-constrained estimators with the conventional local polynomial regression estimators and \citet{he2009double}'s double-smoothing estimator. 

Under the DE-constrained local exponential growth model (\ref{eqn:exp}),  $g^{(k)}(x_0)=\lambda^k g(x_0)$. Table \ref{table:asymptotic} provide a summary comparison of the asymptotic properties of various estimators. The order follows the pattern from the largest to the smallest bias magnitude, approximately. 

The double-smoothed estimator (DS) was proposed by \citet{he2009double}. The related notation is as follows. 
$B(x_0)=(\mu_2^2-\mu_4)/4[g''(x_0)f''(x_0)/f(x_0)+2(g^{(3)}(x_0)f'(x_0)/f(x_0))+g^{(4)}(x_0)] = (\mu_2^2-\mu_4)/4[\lambda^2g(x_0)f''(x_0)/f(x_0)+2(\lambda^3g(x_0)f'(x_0)/f(x_0))+\lambda^4g(x_0)] $.  $V=\int \{(K*K)(v)-(L*L)(v)/\mu_2\}^2dv$, where $L(u)=uK_1(u)$, and  $v_k = \int w^kK_1^2(w)dw$, for $k=0,1,2,...$.

\begin{table}[h!]
  \begin{center}  
    \begin{tabular}{c|c|c} 
      \hline
      \textbf{Method} & \textbf{Asymptotic bias in interior} & \textbf{Asymptotic variance}\\
      \hline
      NW & $\frac{1}{2}\left\{\lambda^2g(x_0)+2\lambda\frac{g(x_0)f'(x_0)}{f(x_0)}\right\}h^2 \mu_2+o_p(h^2)$  & $\frac{1}{nh}\frac{\sigma^2}{f(x_0)}v_0+o_p(\frac{1}{nh})$ \\
      LL & $\frac{1}{2}\lambda^2g(x_0)h^2 \mu_2 +o_p(h^2)$ & $\frac{1}{nh}\frac{\sigma^2(x_0)}{f(x_0)}v_0+o_p(\frac{1}{nh})$\\
      $\text{DE}_1$ & $\frac{1}{2}\lambda^2g(x_0)h^2\mu_2+o_p(h^2)$ & $\frac{1}{nh}\frac{\sigma^2}{f(x_0)}v_0+o_p(\frac{1}{nh})$ \\
      $\text{DE}_2$  & $\frac{1}{6}\lambda^3g(x_0)h^4\mu_4\left(\lambda + \frac{f'(x_0)}{f(x_0)}\right)\mu_4+ o_p(h^4)$ & $\frac{1}{nh}\frac{\sigma^2}{f(x_0)}v_0+o_p(\frac{1}{nh})$\\  
      LQ  & $\frac{1}{24}\frac{\mu_2\mu_6-\mu_4^2}{\mu_2^2-\mu_4}\left\{\lambda^4g(x_0)+4\lambda^3\frac{g(x_0)f'(x_0)}{f(x_0)}\right\}h^4+o_p(h^4)$  & $\frac{1}{nh}\frac{\sigma^2}{f(x_0)}\frac{\mu_4^2v_0-2\mu_2\mu_4v_2+\mu_2^2v_4}{(\mu_2^2-\mu_4)^2}+o_p(\frac{1}{nh})$\\ 
      LC & $ \frac{1}{24}\frac{\mu_2\mu_6-\mu_4^2}{\mu_2^2-\mu_4}\lambda^4g(x_0)h^4+o_p(h^4)$  & $\frac{1}{nh}\frac{\sigma^2}{f(x_0)}\frac{\mu_4^2v_0-2\mu_2\mu_4v_2+\mu_2^2v_4}{(\mu_2^2-\mu_4)^2}+o_p(\frac{1}{nh})$\\ 
      DS & $h^4B(x_0)+o_p(h^4)$ & $\frac{1}{nh}\frac{\sigma^2}{f(x_0)}V+o_p(\frac{1}{nh})$\\       
      $\text{DE}_3$ &$\frac{1}{24}\lambda^4g(x_0)h^4\mu_4+o_p(h^4)$ & $\frac{1}{nh}\frac{\sigma^2}{f(x_0)}v_0+o_p(\frac{1}{nh})$\\
     \hline      
    \end{tabular}    
    \caption{Summary of asymptotic conditional bias and variance for the NW (local constant), LL (local linear), LC (local cubic), DS (double smoothing), $\text{DE}_1$, $\text{DE}_2$, and $\text{DE}_3$ estimators.  Here, $g(x_0) = g(0)\mbox{e}^{\lambda x_0}$.}
\label{table:asymptotic} 
  \end{center}   
\end{table}

We observe that when the differential equation model is correctly specified, the higher-degree DE-constrained estimator exhibits lower bias than conventional local polynomial estimators of the same degree, while their variances remain similar. This suggests a preference for the higher-degree DE-constrained estimator. Furthermore, under model misspecification, the higher-degree DE-constrained estimator continues to outperform its lower-degree counterpart in terms of estimation accuracy. These findings will be demonstrated in the following simulation study.

\section{Simulation Study on Misspecified Model}

In this section, we conduct simulation studies to evaluate the performance of the DE-constrained local polynomial regression estimator under model misspecification. The analysis compares the estimation accuracy of the proposed approach with conventional local polynomial estimators across various degrees of model misspecification and noise levels. 

\vspace{3mm}
\noindent
\textbf{Simulation Setup}

We assess the DE-constrained regression method against conventional local polynomial estimators using 1000 simulated samples of size 100. The data generation encompasses \textcolor{red}{four} scenarios with varying degrees of misspecification:

\begin{itemize}
\item Scenario 1: $g(x) = \mbox{e}^{x - 0.025x^2}$ 
\item Scenario 2: $g(x) = \mbox{e}^{x - 0.1x^2}$ 
\textcolor{red}{\item Scenario 3: $g(x) = \mbox{e}^{x - 0.25x^2}$}
\textcolor{red}{\item Scenario 4: $g(x) = \mbox{e}^{x - 0.5x^2}$}
\end{itemize}

In each scenario, data are drawn uniformly from interval $[0, 1]$ , with additive Gaussian noise of zero mean and standard deviation $\sigma$. The local exponential growth model (\ref{eqn:exp}) assumes $\lambda = 1$  and and simulations are performed at three different noise levels: $\sigma = 0.1, 0.2, \text{and} 0.3$. This design allows comprehensive comparison under increasing model mismatch and noise.

 For each simulated dataset, we apply the following estimators: NW (local constant), LL (local linear), LQ (local quadratic), LC (local cubic), $\text{DE}_1$, $\text{DE}_2$, $\text{DE}_3$, and NLS (the exponential growth model fit by nonlinear least-squares). Bandwidths for each DE-constrained estimator are selected via leave-one-out cross-validation, a procedure that has proven effective for small sample sizes in local polynomial regression contexts. This approach ensures optimal smoothing parameters tailored to each estimator’s structure.

The comparison criterion used in our simulation study is the Integrated Squared Error (ISE), a standard metric in nonparametric statistics for evaluating smoothing estimators. It is defined as the integral of the squared difference between the estimated function $\hat{g}(x)$ and the true function $g(x)$ over the interval $[a,b]$:

\[
\text{ISE} = \int_a^b (\hat{g}(x)-g(x))^2 dx,
\]

which in practice is approximated numerically by summing over a grid of points::

\[
\text{ISE} \approx \sum_{k=1}^K (\hat{g}(x_k)-g(x_k))^2 \Delta x
\]

where $x_1,x_2,...,x_k$ are grid points, $\Delta x = x_{k+1}-x_k$ in the case of equally spaced points, or the trapezoidal rule is applied for unequally spaced grids. We report the average ISE across the 1000 simulated datasets and visualize the distribution of ISE values using boxplots for each estimator. This approach provides a comprehensive assessment of estimator accuracy and variability.

\vspace{3mm}
\noindent
\textbf{Simulation Results}

\begin{figure}
        \centering
        \vspace{-10mm}
        \includegraphics[width=1\textwidth]{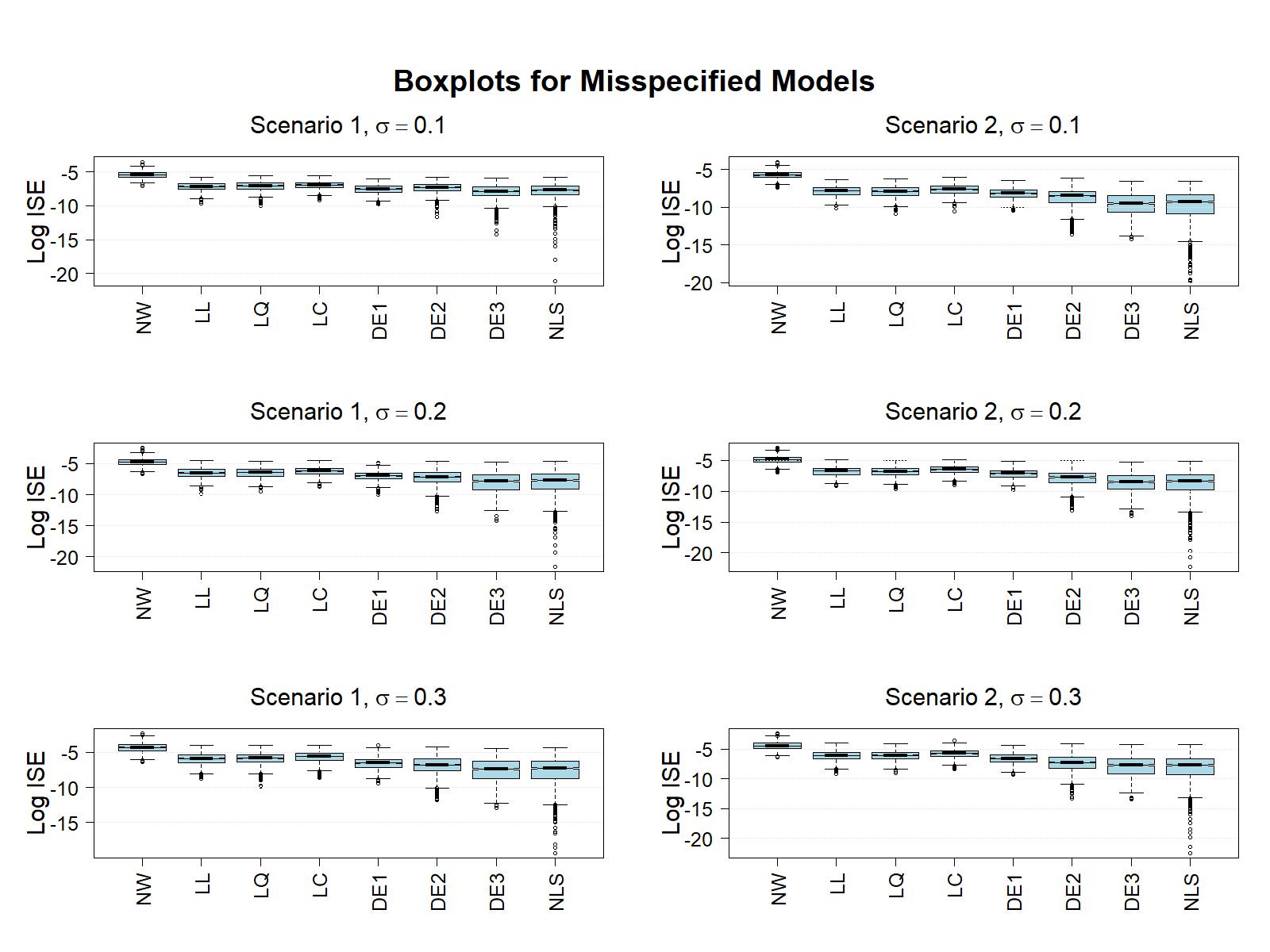}
        \vspace{-9mm}
        \caption{ISE (Integrated Squared Error) distributions (on the log scale) for the estimation methods applied to simulated data with 1000 runs in Scenario 1 and 2. Sample size n = 100. }
        \label{fig:unif}
        \end{figure}

The boxplots in Figure \ref{fig:unif} display the distributions of the Integrated Squared Error (ISE) values for the eight estimation methods applied to the simulated datasets, plotted on a logarithmic scale. Using a logarithmic scale allows clearer visualization and comparison across methods when ISE values span several orders of magnitude. 

In our previous work \citep{ge2025differential}, when the differential equation model was correctly specified, we found that among the DE-constrained estimators for the local growth model, the highest-degree estimator consistently produced the lowest bias. This finding suggests that higher-degree DE-constrained regression is preferable for model selection. Our analysis of boxplots under model misspecification reveals a similar pattern, reinforcing the robustness of the higher-degree estimators. To further illustrate these results, additional figures are provided below, highlighting the improved performance of higher-degree DE-constrained estimators relative to lower-degree counterparts and conventional methods.

\begin{figure}
        \centering
        \vspace{-10mm}
        \includegraphics[width=1\textwidth]{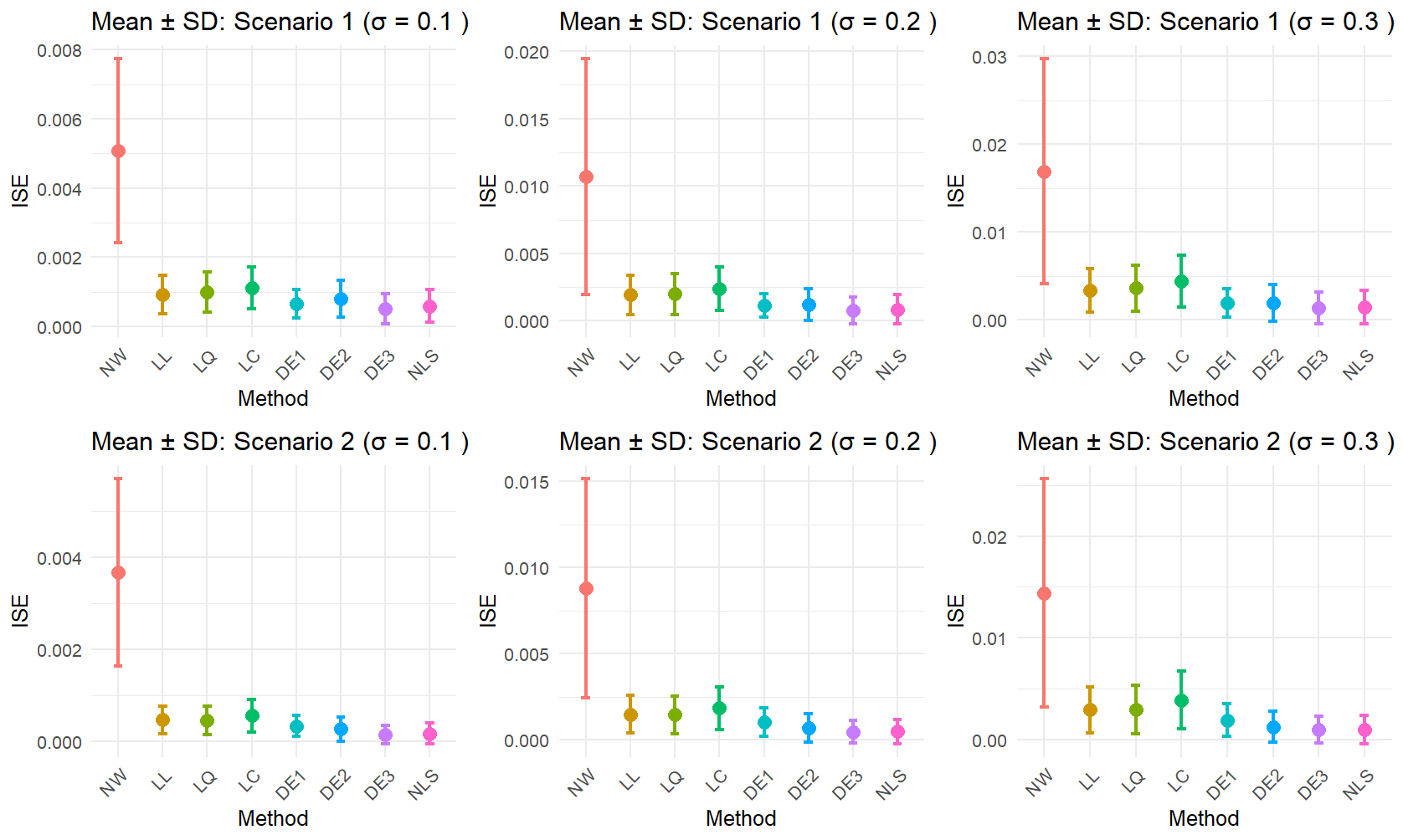}
        \vspace{-9mm}
        \caption{ISE (Integrated Squared Error) mean $\pm$ SD point and error chart for the estimation methods applied to simulated data with 1000 runs in Scenario 1 and 2. Sample size n = 100. }
        \label{fig:meanSD}
        \end{figure}

\begin{figure}
        \centering
        \vspace{-25mm}
        \includegraphics[width=1\textwidth]{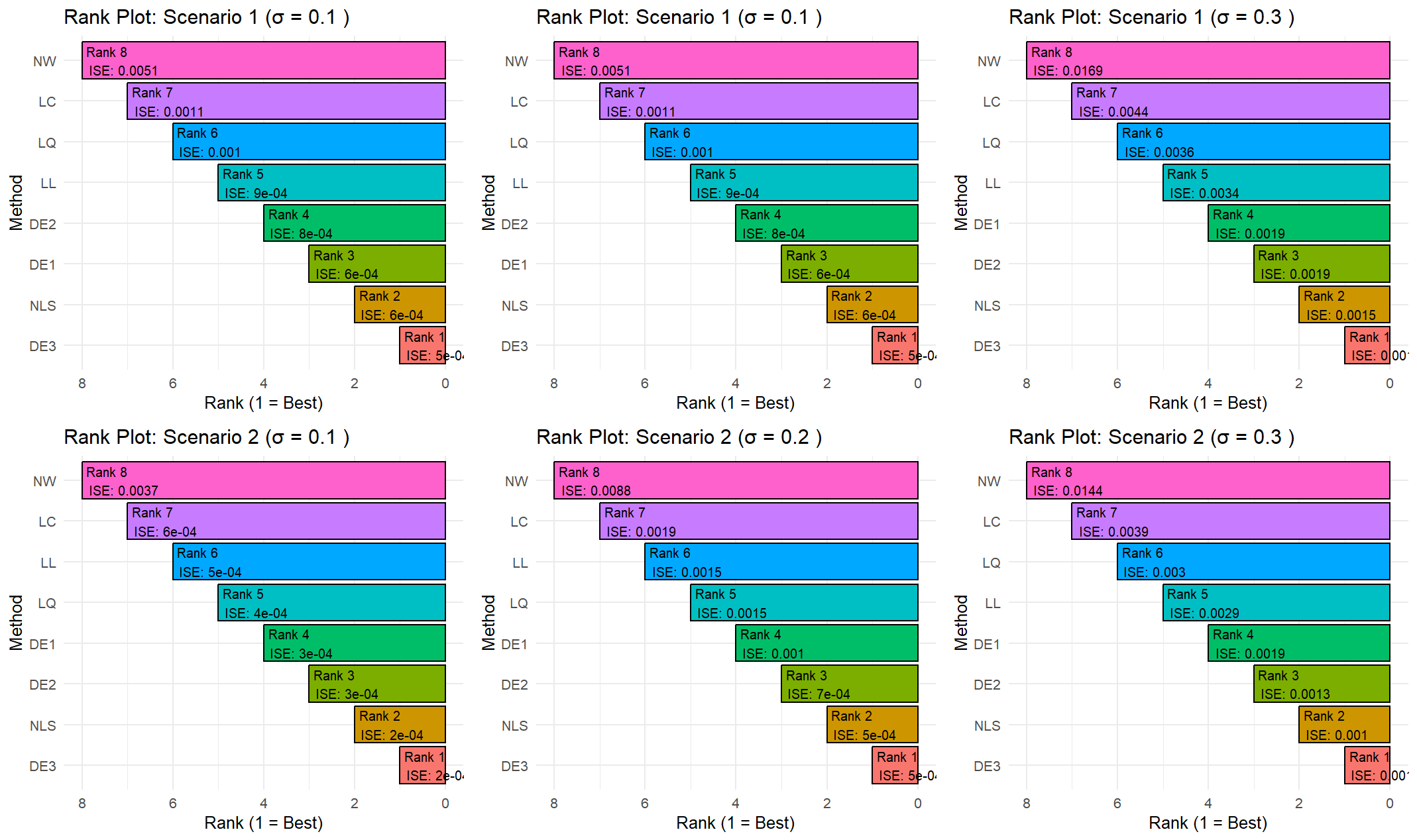}
        \vspace{-9mm}
        \caption{Rank plot for the estimation methods applied to simulated data with 1000 runs in Scenario 1 and 2. Sample size n = 100. }
        \label{fig:rank}
        \end{figure}

\begin{figure}
        \centering
        \vspace{-10mm}
        \includegraphics[width=1\textwidth]{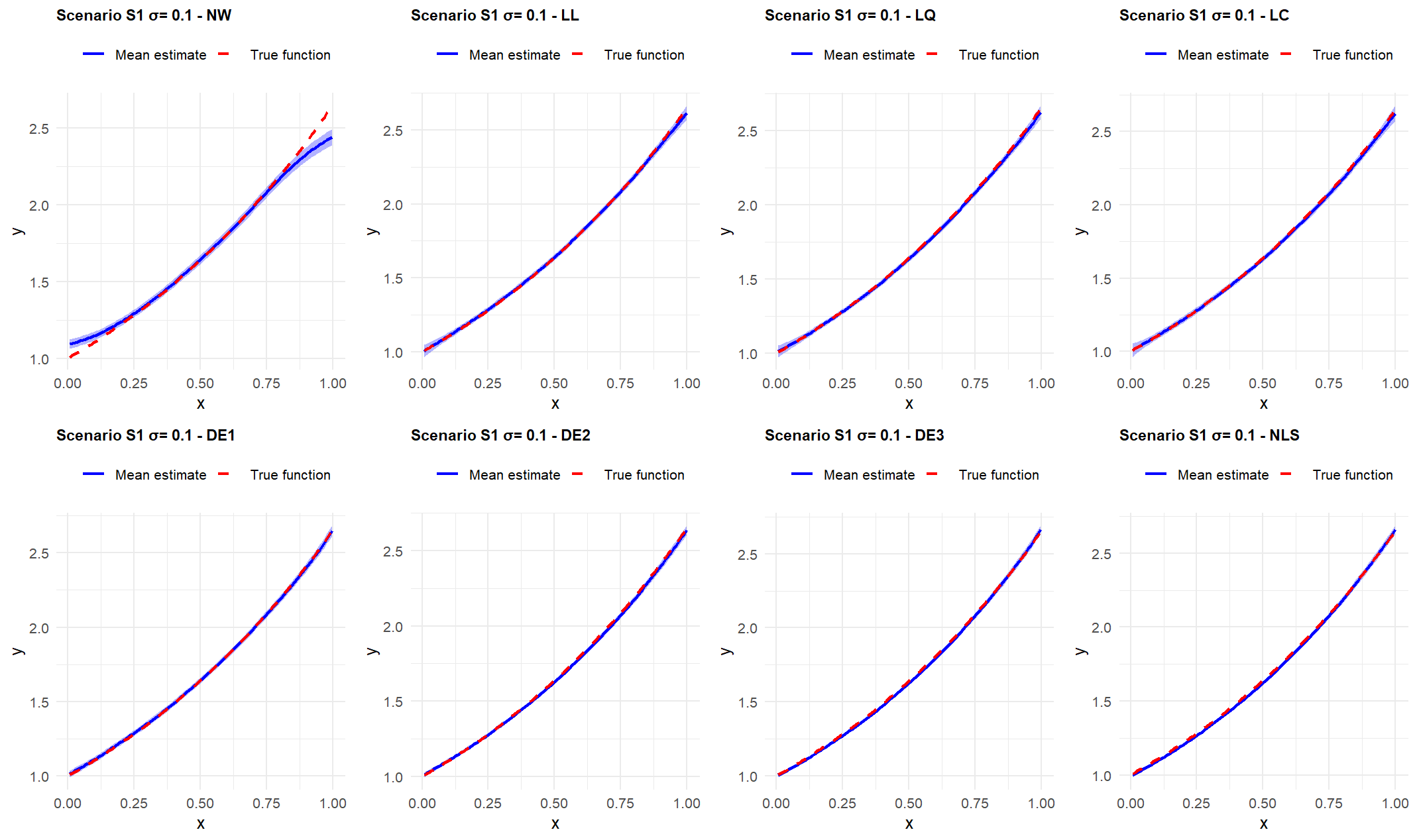}
        \vspace{-9mm}
        \caption{Mean estimated curve $\pm$ SD bands for the estimation methods applied to simulated data with 1000 runs in Scenario 1 ($\sigma=0.1$). Sample size n = 100. }
        \label{fig:curveS1_0.1}
        \end{figure}

Figure \ref{fig:meanSD} presents the mean $\pm$ standard deviation (SD) error bars of the Integrated Squared Error (ISE) across 1000 simulations for both Scenario 1 and Scenario 2. These results demonstrate that the $DE_3$ estimator consistently achieves the best overall performance. Similarly, the rank plot in Figure \ref{fig:rank} supports this conclusion, showing $DE_3$ as the top-ranked method. Figure \ref{fig:curveS1_0.1} illustrates the mean estimated curves with $\pm$ SD bands for Scenario 1 at $\sigma = 0.1$, visualizing how well each method recovers the true underlying function and the stability of estimates. The DE-constrained methods provide superior coverage with narrower SD bands, indicating enhanced accuracy and robustness compared to competitor estimators. Additional curve comparisons for other scenarios are provided in the Appendix.

\section{Discussion}

In Section 5, we discussed the asymptotic bias and variance of the estimators when the model is correctly specified. To further demonstrate the bias reduction capability DE-constrained approach, we use the $DE_1$estimator as an example for comparison with the conventional local constant and local linear estimators under model misspecification.

We suppose that the model we use is the DE-constrained local exponential growth model (\ref{eqn:exp}) with $g'(x)=\lambda_1 g(x)$.

The true function is $g(x)=e^{\lambda_1 x - \lambda_2 x^2}$, where $\lambda_1$ and $\lambda_2$ are known global positive parameters. Its first and second derivatives are given by, 

$$g'(x)=(\lambda_1-2x\lambda_2)g(x),$$

and $$g''(x)=((\lambda_1 - 2x\lambda_2)^2 - 2x \lambda_2)g(x).$$

Using the approach outlined in Section 4, we derive the asymptotic conditional biases of the $DE_1$, NW (local constant), and LL (local linear) estimators under model misspecification. Since the asymptotic conditional variances remain unchanged by misspecification, they are omitted from Table \ref{table:asymptoticmis}. Results confirm that the 
$DE_1$ estimator exhibits the smallest bias among the three methods, demonstrating the bias reduction advantage of the DE-constrained estimator.

\begin{table}[h!]
  \begin{center}  
    \begin{tabular}{c|c}
      \hline
      \textbf{Method} & \textbf{Asymptotic bias in interior}\\
      \hline
      NW & $\frac{1}{2}((\lambda_1 - 2x\lambda_2)^2 - 2x \lambda_2)g(x)h^2 \mu_2+\frac{g'(x)f'(x)}{f(x)}h^2 \mu_2+o_p(h^2)$ \\
      LL & $\frac{1}{2}((\lambda_1 - 2x\lambda_2)^2 - 2x \lambda_2)g(x)h^2 \mu_2+o_p(h^2)$  \\
      DE1-1 & $\frac{1}{2}((\lambda_1 - 2x\lambda_2)^2 - 2x \lambda_2)g(x)h^2 \mu_2 - 2x\lambda_2 g(x)(\lambda_1+\frac{f'(x)}{f(x)})h^2 \mu_2+o_p(h^2)$  \\
      \hline      
    \end{tabular}    
    \caption{Summary of asymptotic conditional bias and variance for the NW (local constant), LL (local linear), and $DE_1$ (first-degree DE-Assisted) estimators for misspecified models.} 
    \label{table:asymptoticmis} 
  \end{center}   
\end{table}

In summary, our study demonstrates that regardless of whether the model is correctly specified or misspecified, fully nonparametric methods are generally outperformed by differential equation (DE)-constrained approaches. The results of this paper and \citet{ge2025differential} demonstrate that the DE-constrained estimator, particularly of higher degree, consistently exhibits reduced bias and improved performance in both correctly specified and misspecified scenarios. 

This finding motivates further research on differential equation–constrained regression models. In our recent work \citep{ge2025local}, we examined a local quasi-exponential model whose structure is governed by a quasi-exponential differential equation. In future work, we plan to extend DE-constrained regression to more general classes of differential equations.

\section*{Acknowledgements}

This research has been supported in part by a grant from the Natural Sciences and Engineering Research Council of Canada (NSERC).

\begin{appendices}

\section*{Appendix}
\subsection*{Proof of Theorems}
\subsubsection{Proof of Theorem 1}
The asymptotic conditional bias of the $k^{th}$-degree estimator $\hat{g}_k (x_0)$ is given by

\begin{equation}
\begin{split}
\mathrm{Bias}(\widehat{g}_k (x_0)|x_1,...,x_n) &=\mathrm{E}[\hat{g}_k(x_0)|x_1,...x_n]-g(x_0) \\
& \approx \frac{1}{(k+1)!}\lambda^{k+1} g(x_0) \frac{\sum_{i=1}^n (x_i-x_0)^{k+1}\{\sum_{p=0}^k \frac{1}{p!}(x_i-x_0)^p \lambda^{p}\}K_h(x_i-x_0) }{\sum_{i=1}^n\{\sum_{p=0}^k \frac{1}{p!}(x_i-x_0)^p \lambda^{p} \}^2K_h(x_i-x_0)}\\
& \approx \frac{1}{(k+1)!}\lambda^{k+1} g(x_0)  \frac{\int_a^b (z-x)^{k+1}\{\sum_{p=0}^k \frac{1}{p!}(z-x)^p \lambda^{p}\}f(z)K_h(z-x)dz}{\int_a^b\{\sum_{p=0}^k \frac{1}{p!}(z-x)^p \lambda^{p} \}^2f(z)K_h(z-x)dz}
\end{split}
\end{equation}
 Let $\frac{z-x}{h}=w$. Then the integral in the numerator becomes
\begin{equation*}
\begin{split}
& \int_a^b (z-x)^{k+1}\left\{\sum_{p=0}^k \frac{1}{p!}(z-x)^p \lambda^{p}\right\}f(z)K_h(z-x)dz \\
&= \int (hw)^{k+1}\left\{\sum_{p=0}^k \frac{1}{p!}(hw)^p \lambda^{p}\right\}f(x+hw)K(w)dw \\
&= \int (hw)^{k+1}\left\{\sum_{p=0}^k \frac{1}{p!}(hw)^p \lambda^{p}\right\}(f(x_0)+hwf'(x_0))K(w)dw \\
&= \int\left \{ \sum_{p=0}^k\frac{1}{p!}h^{p+k+1}w^{p+k+1} \lambda^{p}f(x_0)+\sum_{p=0}^k\frac{1}{p!}h^{p+k+2}w^{p+k+2} \lambda^{p}f'(x_0) \right\}K_1(w)dw. 
\end{split}
\end{equation*}
The denominator can be approximated by
\begin{equation*}
\begin{split}
& \int_a^b\left\{\sum_{p=0}^k \frac{1}{p!}(z-x)^p \lambda^{p}\right \}^2f(z)K_h(z-x)dz \\
& = \int \left\{\sum_{p=0}^k \frac{1}{p!}(hw)^p \lambda^{p} \right\}^2f(x+hw)K_1(w)dw \\
& \approx f(x_0)
\end{split}
\end{equation*}
when $x_0 \in (a+h, b-h)$, and when $k$ is odd, the asymptotic conditional bias of $\widehat{g}_k (x_0)$ is
\begin{equation}
\mathrm{Bias}(\widehat{g}_k (x_0)|x_1,...,x_n) = \frac{1}{(k+1)!}\lambda^{k+1}g(x_0)h^{k+1}\mu_{k+1}+o_p(h^{k+1})
\end{equation}
and when $k$ is even, 
\begin{equation}
\mathrm{Bias}(\widehat{g}_k (x_0)|x_1,...,x_n) = \frac{1}{(k+1)!}\lambda^{k+1}g(x_0)h^{k+2}\mu_{k+2}(\frac{\lambda}{k+2}+\frac{f'(x_0)}{f(x_0)})+o_p(h^{k+2}). 
\end{equation}

\subsubsection{Proof of Theorem 2}

Assume $x_0 \in (a+h, b-h)$.  
The asymptotic conditional variance of the $k^{th}$-degree estimator (DE1-$k$ estimator) is then
\begin{align}
\mathrm{Var}(\widehat{g}_k (x_0)|x_1,...,x_n)
&= \frac{\sum_{i=1}^n  \mathrm{Var}(y_i|x_1,...,x_n) \{\sum_{p=0}^k \frac{1}{p!}(x_i-x_0)^p \lambda^{p} \}^2 K_h^2(x_i-x_0) }{ \{\sum_{i=1}^n \{\sum_{p=0}^k \frac{1}{p!}(x_i-x_0)^p \lambda^{p} \}^2K_h(x_i-x_0) \}^2} \nonumber \\
&= \frac{\sum_{i=1}^n  \sigma^2 \{\sum_{p=0}^k \frac{1}{p!}(x_i-x_0)^p \lambda^{p} \}^2 K_h^2(x_i-x_0) }{ \{\sum_{i=1}^n \{\sum_{p=0}^k \frac{1}{p!}(x_i-x_0)^p \lambda^{p} \}^2K_h(x_i-x_0) \}^2} \nonumber \\
& \approx \frac{1}{n} \frac{\int_a^b  \sigma^2 \{1+\sum_{p=1}^k \frac{1}{p!}(z-x)^p \lambda^{p} \}^2 K_h^2(z-x)f(z)dz }{ \{\int_a^b \{1+\sum_{p=1}^k \frac{1}{p!}(z-x)^p \lambda^{p} \}^2K_h(z-x) f(z)dz\}^2} \nonumber \\
&=\frac{1}{nh} \frac{\int  \sigma^2 \{1+\sum_{p=1}^k \frac{1}{p!}(hw)^p \lambda^{p} \}^2 K^2(w)f(x+hw)dw }{ \{\int \{1+\sum_{p=1}^k \frac{1}{p!}(hw)^p \lambda^{p} \}^2K(w) f(x+hw)dw\}^2} \nonumber \\
&\approx \frac{\sigma^2R(K)}{nhf(x_0)}+o_p\left(\frac{1}{nh}\right)
\end{align}
where $\frac{z-x}{h}=w$.
If $\sigma^2$ varies with  $x$, that is, the conditional variance $\mathrm{Var}(y_i|x_1,...,x_n) = \sigma^2(x_i)$, then the asymptotic conditional variance of $\widehat{g}_k (x_0)|x_1,...,x_n)$
is given by
\begin{align}
\mathrm{Var}(\widehat{g}_k (x_0)|x_1,...,x_n)
&= \frac{\sum_{i=1}^n  \mathrm{Var}(y_i|x_1,...,x_n) \{\sum_{p=0}^k \frac{1}{p!}(x_i-x_0)^p \lambda^{p} \}^2 K_h^2(x_i-x_0) }{ \{\sum_{i=1}^n \{\sum_{p=0}^k \frac{1}{p!}(x_i-x_0)^p \lambda^{p} \}^2K_h(x_i-x_0) \}^2} \nonumber \\
&= \frac{\sum_{i=1}^n  \sigma^2(x_i) \{\sum_{p=0}^k \frac{1}{p!}(x_i-x_0)^p \lambda^{p} \}^2 K_h^2(x_i-x_0) }{ \{\sum_{i=1}^n \{\sum_{p=0}^k \frac{1}{p!}(x_i-x_0)^p \lambda^{p} \}^2K_h(x_i-x_0) \}^2}. \nonumber
\end{align}
Applying the first-degree Taylor expansion for $\sigma^2(x_i)$ in a sufficiently small neighborhood of  $x_0$, we have 
$$\sigma^2(x_i) \approx \sigma^2(x_0) + (x_i-x_0) \frac{d\sigma^2(x_0)}{dx} \approx \sigma^2(x_0)$$
since the derivative $\frac{d\sigma^2(x_0)}{dx}$ is bounded on $[a,b]$, by Assumption (IV).  
Then we can obtain 
\begin{align}
\mathrm{Var}(\widehat{g}_k (x_0)|x_1,...,x_n)
& \approx \frac{1}{n} \frac{\int_a^b  \sigma^2(x_0) \{1+\sum_{p=1}^k \frac{1}{p!}(z-x)^p \lambda^{p} \}^2 K_h^2(z-x)f(z)dz }{ \{\int_a^b \{1+\sum_{p=1}^k \frac{1}{p!}(z-x)^p \lambda^{p} \}^2K_h(z-x) f(z)dz\}^2} \nonumber \\
&=\frac{1}{nh} \frac{\sigma^2(x_0)\int  \{1+\sum_{p=1}^k \frac{1}{p!}(hw)^p \lambda^{p} \}^2 K^2(w)f(x+hw)dw }{ \{\int \{1+\sum_{p=1}^k \frac{1}{p!}(hw)^p \lambda^{p} \}^2K(w) f(x+hw)dw\}^2} \nonumber \\
&\approx \frac{\sigma^2(x_0)R(K)}{nhf(x_0)}+o_p\left(\frac{1}{nh}\right)
\end{align}


        \begin{figure}
        \centering
        \vspace{-10mm}
        \includegraphics[width=1\textwidth]{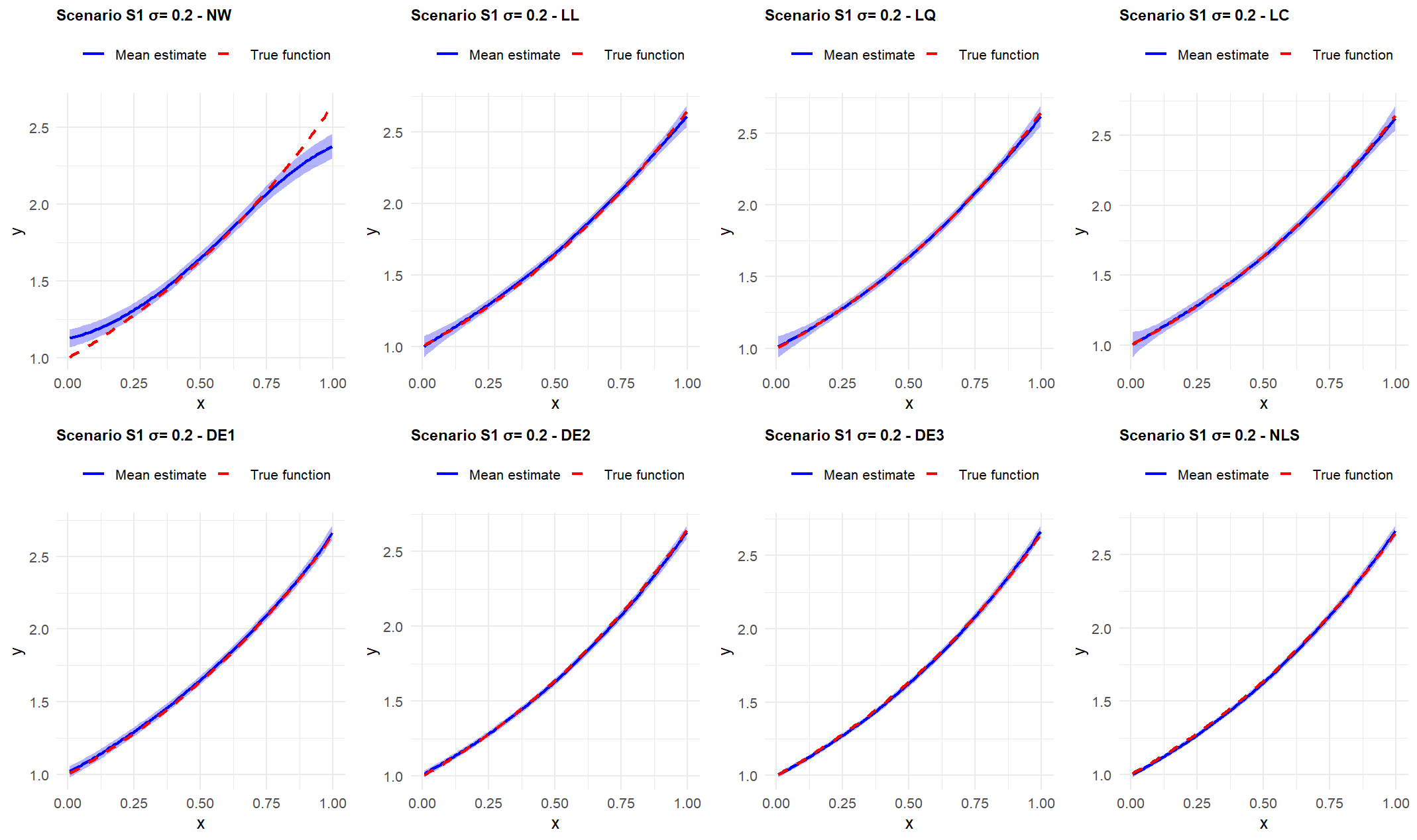}
        \vspace{-9mm}
        \caption{Mean estimated curve $\pm$ SD bands for the estimation methods applied to simulated data with 1000 runs in Scenario 1 ($\sigma=0.2$). Sample size n = 100. }
        \label{fig:curveS1_0.2}
        \end{figure}

        \begin{figure}
        \centering
        \vspace{-10mm}
        \includegraphics[width=1\textwidth]{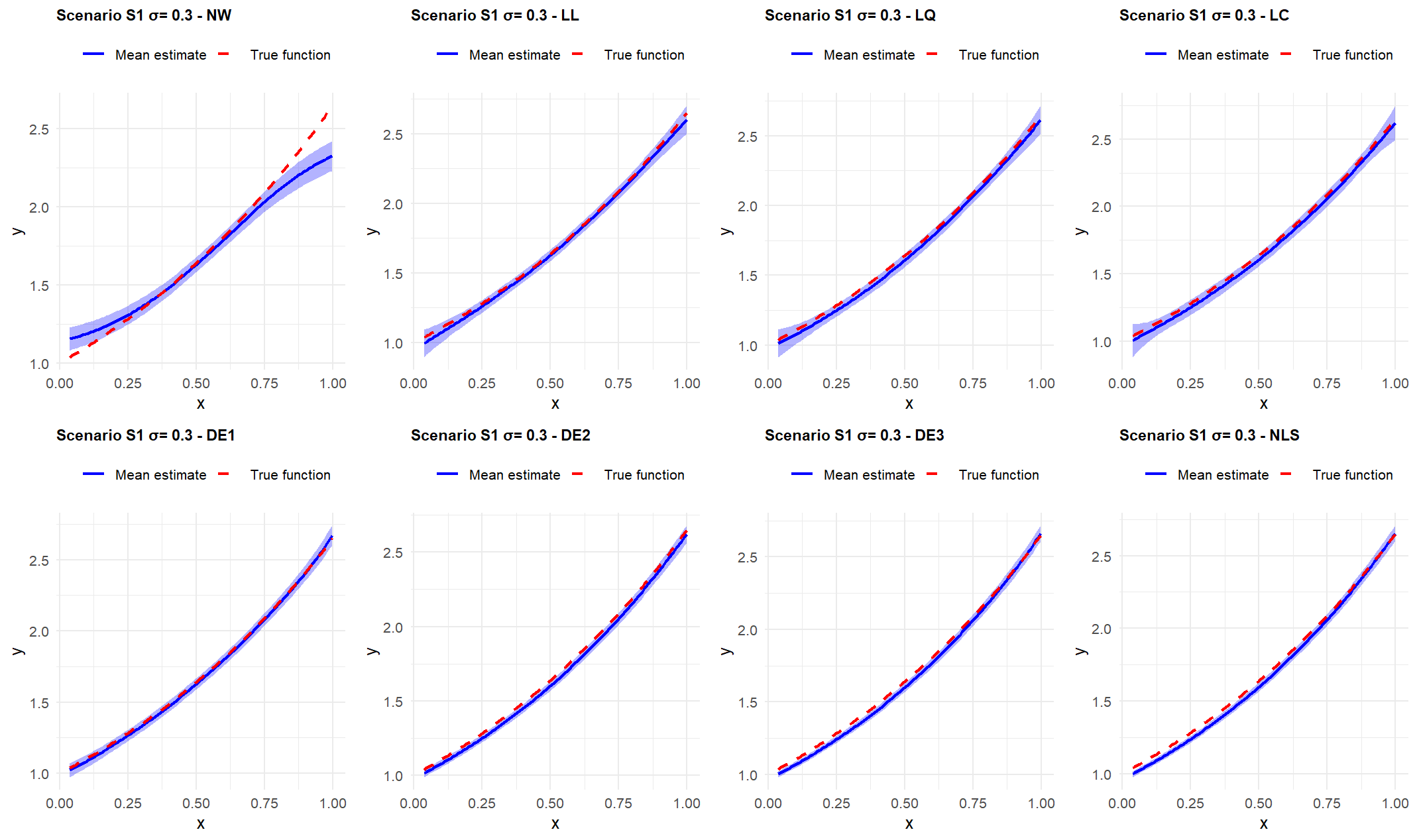}
        \vspace{-9mm}
        \caption{Mean estimated curve $\pm$ SD bands for the estimation methods applied to simulated data with 1000 runs in Scenario 1 ($\sigma=0.3$). Sample size n = 100. }
        \label{fig:curveS1_0.3}
        \end{figure}

        \begin{figure}
        \centering
        \vspace{-10mm}
        \includegraphics[width=1\textwidth]{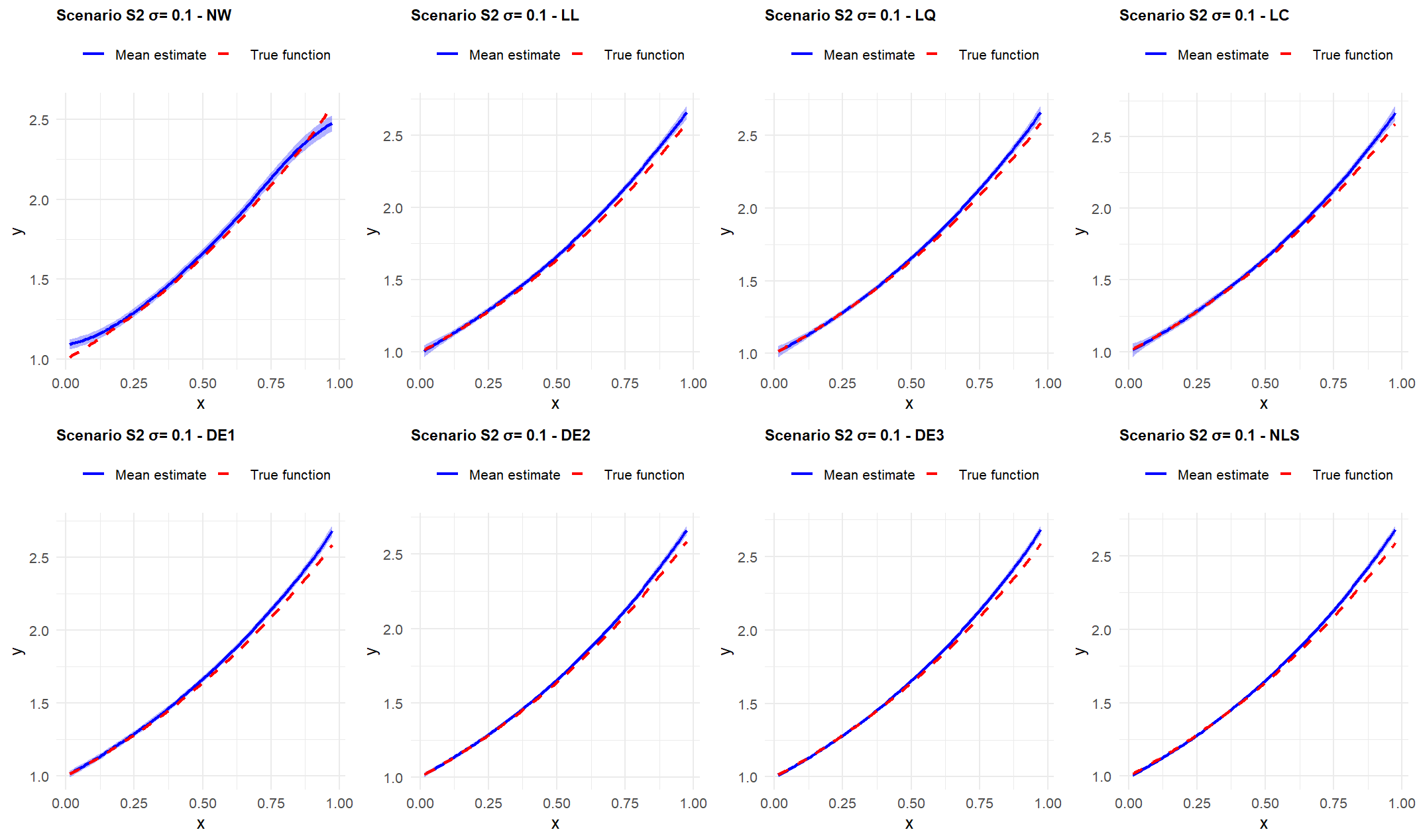}
        \vspace{-9mm}
        \caption{Mean estimated curve $\pm$ SD bands for the estimation methods applied to simulated data with 1000 runs in Scenario 2 ($\sigma=0.1$). Sample size n = 100. }
        \label{fig:curveS2_0.1}
        \end{figure}

        \begin{figure}
        \centering
        \vspace{-10mm}
        \includegraphics[width=1\textwidth]{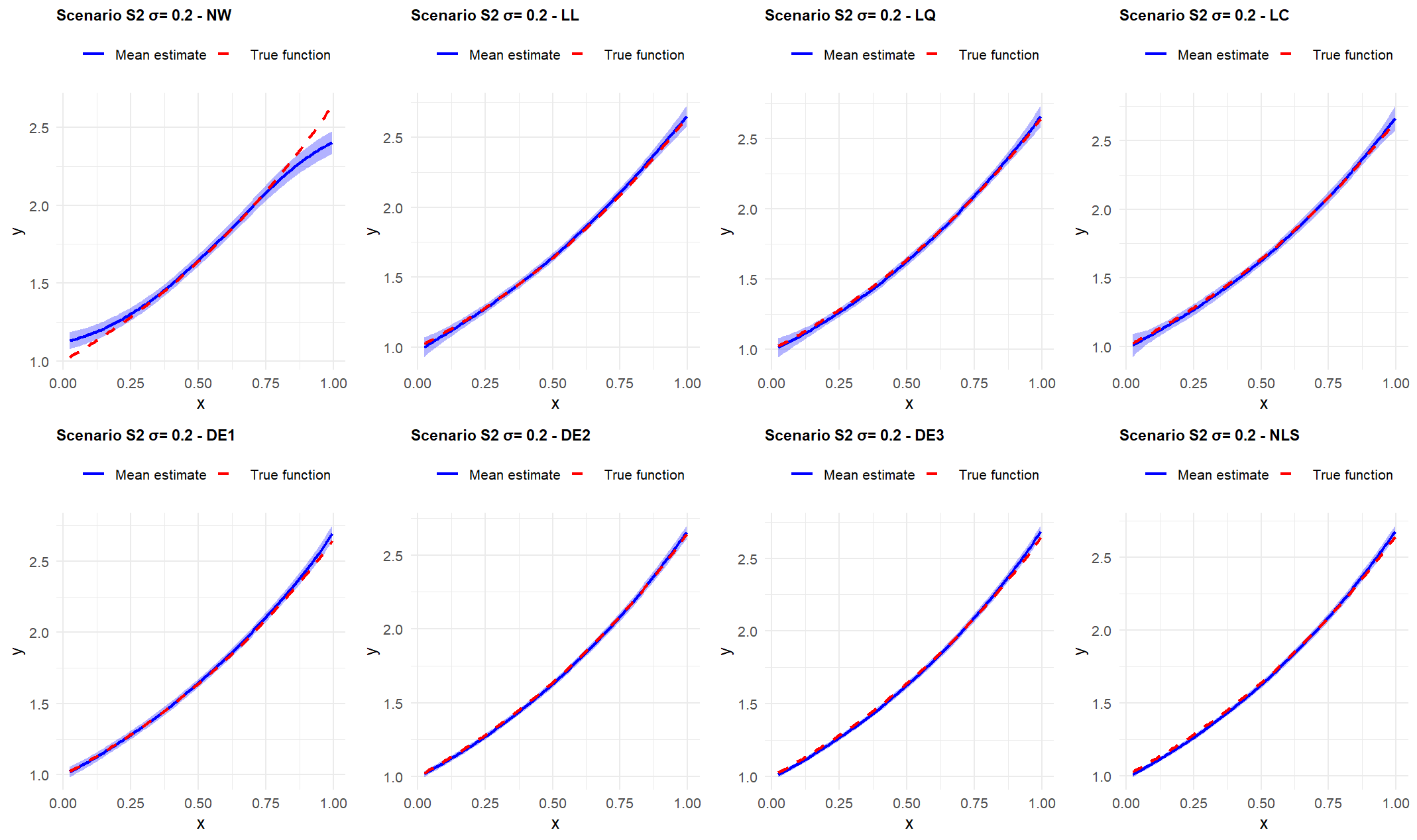}
        \vspace{-9mm}
        \caption{Mean estimated curve $\pm$ SD bands for the estimation methods applied to simulated data with 1000 runs in Scenario 2 ($\sigma=0.2$). Sample size n = 100. }
        \label{fig:curveS2_0.2}
        \end{figure}

        \begin{figure}
        \centering
        \vspace{-10mm}
        \includegraphics[width=1\textwidth]{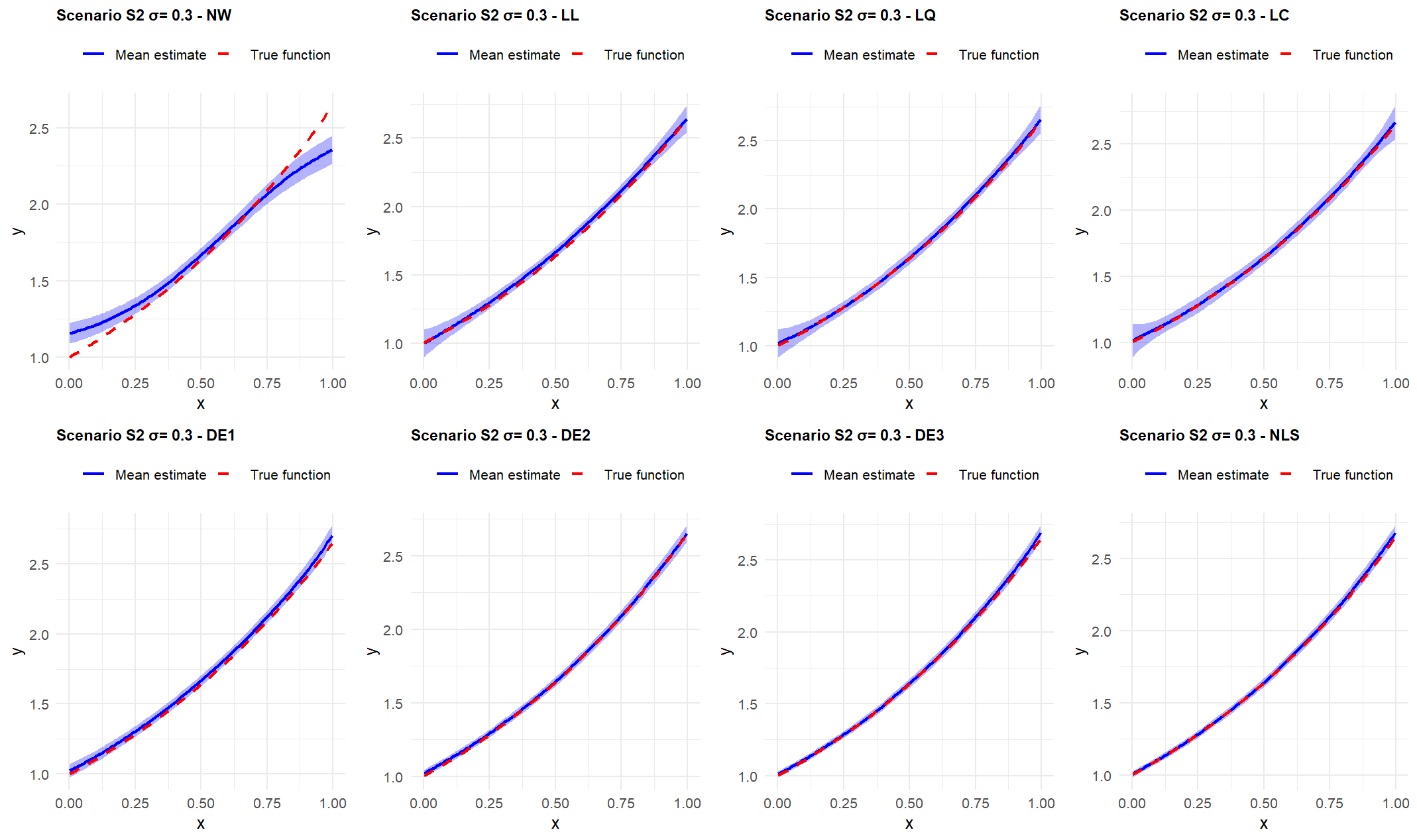}
        \vspace{-9mm}
        \caption{Mean estimated curve $\pm$ SD bands for the estimation methods applied to simulated data with 1000 runs in Scenario 2 ($\sigma=0.3$). Sample size n = 100. }
        \label{fig:curveS2_0.3}
        \end{figure}

\end{appendices}

\end{document}